\documentclass[twocolumn]{aastex63}

\usepackage{tabularx}
\usepackage{url}
\usepackage{graphicx}
\usepackage[T1]{fontenc}
\usepackage{ae,aecompl}
\usepackage{booktabs}
\usepackage{natbib}

\usepackage{blindtext}
\usepackage{longtable}
\usepackage{tabu}
\usepackage{lineno}

\def\unit #1{\,{\rm #1}}

\newcommand\kms{\rm \,\unit{km\,s^{-1}}}

\newcommand\cmsqi{\rm \,\unit{cm^{-2}}}

\newcommand\cm{\rm \,\unit{cm}}

\newcommand\cmcubei{\rm \,\unit{cm^{-3}}}

\newcommand\kev{\rm \,\unit{keV}}

\newcommand\lunit{\rm \,erg \,s^{-1}}

\newcommand\xiunit{\rm \,erg\,cm\,s^{-1}}

\newcommand\ledd{L_{\rm \, Edd}}
\newcommand\lion{L_{\rm \, ion}}

\newcommand\lambdaedd{\lambda_{\rm \, Edd}}

\newcommand\lbol{L_{\rm \, bol}}

\newcommand\msol{M_{\odot}}

\newcommand\mbh{M_{\rm BH}}

\newcommand\msolyi{M_{\odot}\,\rm yr^{-1}}

\newcommand\nh{ N_{\rm H}}

\newcommand\pc{\unit{pc}}

\newcommand\kpc{\unit{kpc}}

\newcommand\AAA{\,\rm \mathring{A}}

\newcommand\ev{\unit{\, eV}}

\newcommand\vout{v_{\rm out}}

\newcommand\mout{\dot{M}_{\rm out}}

\newcommand\pout{\dot{P}_{\rm out}}

\newcommand\pbol{\dot{P}_{\rm bol}}
\newcommand\pwind{\dot{P}_{\rm w}}

\newcommand\Ek{\dot{E}_{\rm K}}

\newcommand\hst{{\it Hubble Space Telescope}}
\newcommand\fuse{{\it Far Ultraviolet Spectroscopic Explorer}}
\newcommand\COS{{\it Cosmic Origins Spectrograph}}

\newcommand\chandra{{\it Chandra}}
\newcommand\suzaku{{\it Suzaku}}

\newcommand\nustar{{\it NuSTAR}}
\newcommand\xmm{{\it XMM-Newton}}
\newcommand\athena{{\it Athena}}
\newcommand\arcus{{\it Arcus}}
\newcommand\xrism{{\it XRISM}}
\newcommand\lynx{{\it Lynx}}
\newcommand\luvoir{{\it LUVOIR}}
\newcommand\eht{{\it Event Horizon Telescope}}

\def\aap{A\&A} \def\apjl{ApJ}  
 \def\apjs{ApJS}

\published{review in \textbf{\large Nature Astronomy}}

\shorttitle{Ionized outflows from active galactic nuclei as the essential elements of feedback}

\begin{document}

\title{ Ionized outflows from active galactic nuclei as the essential elements of feedback}

\author[0000-0003-2714-0487]{Sibasish Laha} 
\affiliation{Astroparticle physics laboratory, NASA Goddard Space Flight Center,Greenbelt, MD 20771, USA.}
\affiliation{Center for Space Science and Technology, University of Maryland Baltimore County, 1000 Hilltop Circle, Baltimore, MD 21250, USA.}

\author[0000-0003-2714-0487]{Christopher S. Reynolds} 
\affiliation{Institute of Astronomy, University of Cambridge, Madingley Road, Cambridge CB3 0HA, UK.}

\author[0000-0003-2714-0487]{James Reeves} 
\affiliation{Department of Physics, Institute for Astrophysics and Computational Sciences, The Catholic University of America, Washington, DC 20064, USA.}
\affiliation{INAF, Osservatorio Astronomico di Brera, Via Bianchi 46 I-23807 Merate (LC), Italy.}

\author[0000-0003-2714-0487]{Gerard Kriss} 
\affiliation{Space Telescope Science Institute, 3700 San Martin Drive, Baltimore, MD 21218, USA.}

\author[0000-0000-0000-0000]{Matteo Guainazzi} 
\affiliation{ESTEC/ESA, Keplerlaan 1, 2201AZ Noordwijk, The Netherlands.}

\author[0000-0003-2714-0487]{Randall Smith} 
\affiliation{High Energy Astrophysics, Center for Astrophysics | Harvard \& Smithsonian, Cambridge, MA, USA.}

\author[0000-0003-2714-0487]{Sylvain Veilleux} 
\affiliation{Department of Astronomy, University of Maryland, College Park, MD 20742, USA.}
\affiliation{Joint Space-Science Institute, University of Maryland, College Park, MD 20742, USA.}

\author[0000-0003-2714-0487]{Daniel Proga} 
\affiliation{Department of Physics \& Astronomy, University of Nevada, Las Vegas, USA.}

\correspondingauthor{Sibasish Laha}
\email{sibasish.laha@nasa.gov, sib.laha@gmail.com}



\begin{abstract}
	
Outflows from active galactic nuclei (AGN) are one of the fundamental mechanisms by which the central supermassive black hole interacts with its host galaxy. Detected in $\ge 50\%$ of nearby AGN, these outflows have been found to carry kinetic energy that is a significant fraction of AGN power, and thereby give `negative’ feedback to their host galaxies. To understand the physical processes that regulate them, it is important to have a robust estimate of their physical and dynamical parameters. In this review we summarize our current understanding on the physics of the ionized outflows detected in absorption in the UV and X-ray wavelength bands. We discuss the most relevant observations and our current knowledge and uncertainties in the measurements of the outflow parameters. We also discuss their origin and acceleration mechanisms. The commissioning and concept studies of large telescope missions with high resolution spectrographs in UV/optical and X-rays along with rapid advancements in simulations offer great promise for discoveries in this field over the next decade.

\end{abstract}


\vspace{0.5cm}

\section{Introduction}

 Active Galactic Nuclei (AGN) are the power houses at the centre of active galaxies hosting a super massive black hole (SMBH). It is now widely accepted that the central SMBH accretes matter via an accretion disk, resulting in a wide range of observable AGN types such as low-luminosity AGN, Seyferts, Quasars and radio galaxies \citep{2019A&A...630A..94G}. The accretion disk radiates a significant amount of the accretion power (up to $10^{47}\lunit$ in some AGN) and produces outflows in the forms of winds and/or relativistic jets. Much have been theorized and observed in this topic in the last four decades, but, several fundamental questions still elude us. For example, (1) what is the exact geometry of the central engine and its surroundings? (2) How does matter at large distances ($\ge \pc$) lose angular momentum, fall in and feed the accretion disk? (3) How are powerful outflows generated from the heart of these massive sources? and (4) How do the radiation and particle outflows couple with the host galaxy gas, the one very important factor that determines the effectiveness of how an AGN interacts with its host galaxy.

 Observational evidence has clearly pointed out that the outflows from SMBH play a crucial role not only in determining the AGN structure, but also in controlling the AGN and star formation activity \citep[see for e.g., the reviews by][]{2004cbhg.symp..374B,2012ARA&A..50..455F,2017NatAs...1E.165H,2020A&ARv..28....2V}. Hence, AGN feedback has the potential to unravel some of the currently outstanding puzzles in astronomy, such as the physical basis behind the strong correlation between SMBH mass and host galaxy stellar bulge velocity dispersion \citep[$\rm \mbh-\sigma$ relation, ][]{2001ApJ...547..140M}, the cooling flow problem in intra-cluster medium (ICM), and the evolution of the quasar luminosity function across cosmic timescales in different wavelength bands \citep{2005A&A...441..417H,2014ApJ...786..104U,2017MNRAS.466.1160M,2019MNRAS.488.1035K,2020MNRAS.495.3252S}. Ideally speaking, the detection and characterization of the inflow and outflow of gas in the vicinity of the SMBH can answer several of these questions. However, the lack of spatial resolution in the inner $pc$ of the central engines even for the nearest sources (barring a few recent observations by the \eht{} and {\it GRAVITY} telescopes) impedes us from viewing these flows directly. Hence, absorption spectroscopy serves as a vital probe.

 The outflows from AGN primarily act in two modes. The kinetic or maintenance mode, which involves powerful radio jets from the AGN which tend to operate at lower Eddington ratio. The Eddington ratio $\lambda_{\rm Edd}$ is defined as $\lambda_{\rm Edd}=\lbol/\ledd$, where $\lbol$ and $\ledd$ are the bolometric and Eddington luminosity of the AGN respectively, corresponding to the SMBH mass. The other is the radiative or wind mode, which involves non-collimated ionized winds prevalent in AGNs with a range in Eddington ratios \citep[i.e., $\lambda_{\rm Edd}\sim 10^{-2}-1$, see for e.g., ][]{2012ARA&A..50..455F}. This review focuses on the study of the radiative mode of outflows which are mostly detected as ionized blue-shifted absorption lines in UV and X-ray spectra of almost $\sim 50-65\%$ of AGN in the nearby Universe \citep{1995MNRAS.273.1167R,2003ARA&A..41..117C,2007MNRAS.379.1359M,2014MNRAS.441.2613L}. The absorption lines arise from different ionic states of astronomically abundant elements such as H, He, O, N, C, Ne, Fe etc.

To understand the physics and impact of these outflows we need to have an estimate of their physical and dynamical parameters. From spectral fitting of absorption lines/edges in UV-optical and X-rays one obtains the ionization parameter ($\xi$), the column density ($\nh$) and the outflow velocity ($v$). The ionization parameter is defined as $\xi=\lion/(n_{\rm e}r^2)$, where $\lion$ is the ionizing luminosity of the source facing the cloud (integrated over $13.6\ev- 100\kev$), $n_{\rm e}$ is the electron density of the plasma and $r$ is the distance of the cloud from the ionizing source. Assuming a spherical outflow \citep{2005A&A...431..111B}, the dynamics of the outflows can be quantified using three parameters: the mass outflow rate $\mout=1.23 r^2 \,m_p \, n(r) \, \vout \, C_{v}(r) \, C_g$, the momentum outflow rate $\pout=\dot{M}_{\rm out} \, \vout$ and the kinetic luminosity $\Ek=\frac{1}{2} \dot{M}_{\rm out} \, \vout^2$, where the factor $1.23$ accounts for the cosmic abundance of elements while estimating the mass of the outflow, $m_{p}$ is the mass of proton, $n(r)$ is the density of the outflow at radius $r$, $C_{v}(r)$ is the volume filling factor as a function of distance and $C_g$ is the global covering factor. We note that all of these quantities explicity depend on the specific choice of the covering factor, volume filling factor and the distance $r$. In absorption spectroscopy all of these quantities are not straightforward to measure. Hence, special methods and indirect ways are used to estimate them, which we discuss in Section \ref{Sec:phenomeno}. It has been conjectured that these outflows can remove gas from the host galaxies and quench the star formation and AGN activity if their kinetic luminosity is typically $\sim 0.5\%$ of the Eddington limit of the AGN \citep{2010MNRAS.401....7H,2012MNRAS.425..605F,2019MNRAS.484.1829Z}.

In the UV, strong absorption features are common in rest-frame spectra of AGNs ($\sim 500-2000 \AAA$) arising out of several resonant transitions, such as the HI Lyman$\alpha$ lines ($\lambda1216$) and higher ionization doublet states, such as CIV $\lambda \lambda 1549,1551$, where the wavelengths are expressed in $\rm \AAA$. The most common diagnostic absorption lines in UV arise from the ions CIV, NV, OVI, SiIV, CII, SiII and FeII. Similarly in the soft X-rays ($0.2-2\kev$), several ions imprint their absorption signatures and are popularly known as the warm absorbers (WA) and mostly have higher ionization states compared to their UV counterparts \citep[see for e.g.,][ and references therein]{2004ApJ...611...68K,2017A&A...601A..17B,2002ApJ...574..643K,2013MNRAS.430.2650L,2014MNRAS.441.2613L}. Highly-ionized outflows with relativistic velocities $v\sim 0.1c$ (where c is the speed of light), have been detected in the X-ray spectra as Lyman-$\alpha$ and/or He-$\alpha$ resonance absorption lines of Fe at energies $\sim 6.70$ and $\sim 6.96\kev$ respectively, are popularly known as the ultra-fast outflows (UFOs) and are one of the most energetic amongst all the types of ionized outflows detected in quasars \citep{2002ApJ...579..169C,2003ApJ...595...85C,2003MNRAS.346.1025P,2005A&A...442..461D,2006ApJ...646..783M,2007ApJ...670..978B,2009A&A...504..401C,2010A&A...521A..57T,2011A&A...536A..49G,2011MNRAS.414.3307G,2011MNRAS.414.1965L,2012MNRAS.422.1914D}.

Although we make a distinction between the outflows in terms of their different wavelength ranges, we must note that the properties, nature and origin of these outflows may not be different from each other and may share spatial/temporal commonality \citep{2013MNRAS.430.1102T}. Thus, although we discuss their observations separately, we address their possible connections in Section \ref{subsec:connection}.

The review begins with our current observational understanding on the physics of the ionized outflows in Section \ref{Sec:phenomeno}. We discuss the main science questions related to the origin and acceleration mechanisms of the outflows in Section \ref{Sec:origin}. Section \ref{Sec:outstanding_questions}  discusses the outstanding questions in this field of study followed by Section \ref{Sec:instruments}  which lists the future instruments and improved theories required to better our understanding. In a short review of a mature field of AGN ionized outflows it is not possible to cover all aspects of the subject and some subjectivity may creep in with unintended bias.

\begin{figure*}
  \centering

	\includegraphics[width=12cm,angle=0]{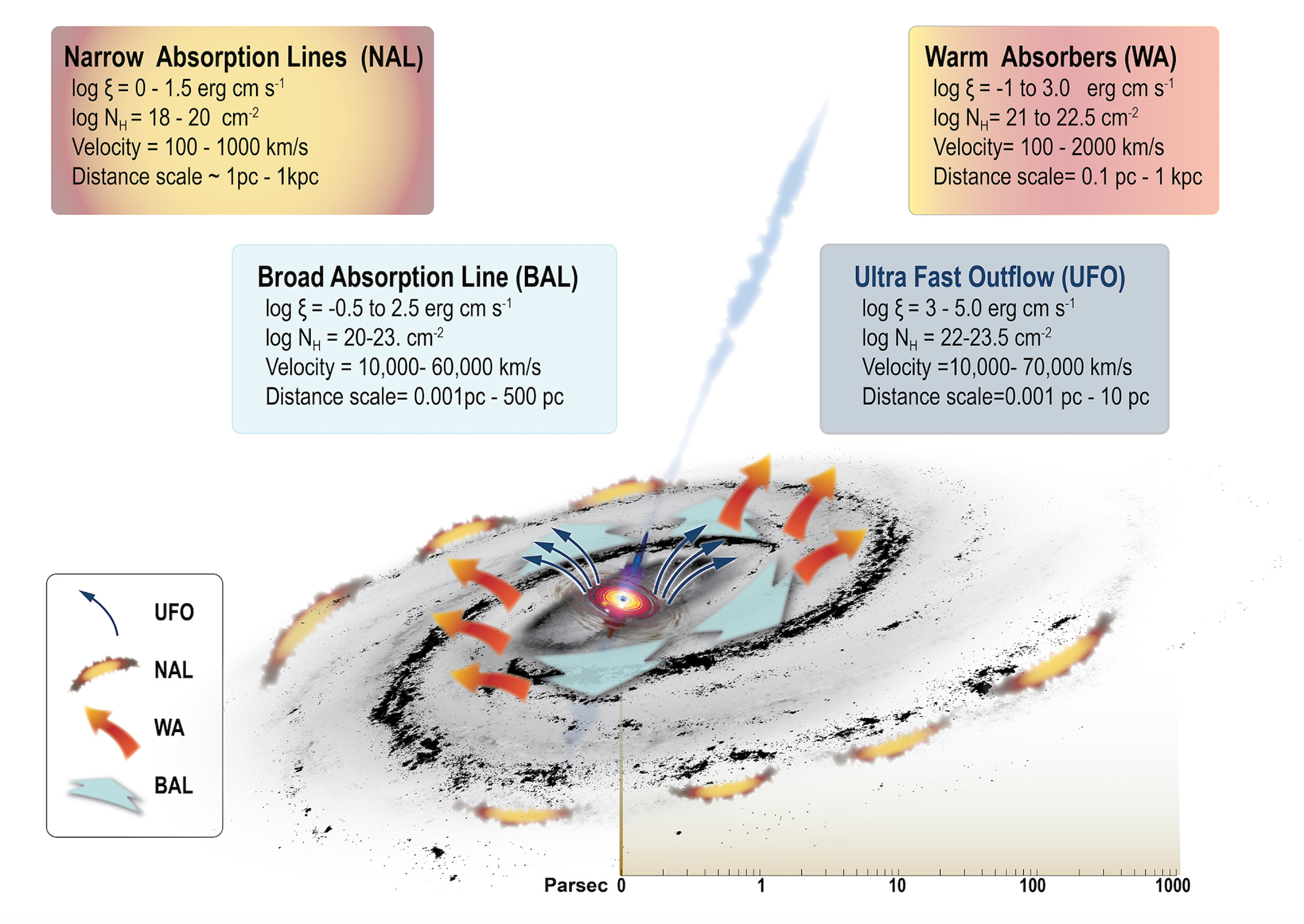}
	
	\caption{ The different ionized outflows detected in AGN and their average physical parameters. We note that there is considerable diversity in the properties of each type of outflows depending on the nature of the central AGN. In addition, there is substantial overlap in parameter space of the different outflows, including their distance scales.} \label{Fig:outflows}
\end{figure*}


\section{Phenomenology of the ionized outflows }\label{Sec:phenomeno}

The most significant advancement in our knowledge of ionized outflows in AGN happened after the advent of high spectral and spatial resolution observatories. The \hst{} (launched in 1990), and \fuse{} (launched in 1999) started an era of high resolution optical/UV spectroscopic studies of AGN outflows. The far UV instrument \COS{} which was put on board the \hst{} in 2009 specialized in spectroscopy of faint sources with a resolving power of $\sim 1500- 20,000$, and hence added immensely to our knowledge. The revolution in X-ray spectroscopy came about with the launch of \xmm{} and \chandra{} in 1999, both of which have high resolution grating X-ray spectrometres, yielding a resolution of up to $\sim 800$ in the soft X-ray energy band. The broad band $0.3-10 \kev$ spectra of the EPIC-pn onboard \xmm{} with an average spectral resolution of $\sim 100$ yielded unprecedented detections of UFOs in the $\sim 6.4-9\kev$ energy range. More recently \suzaku{} (2005-2015) and \nustar{} (launched 2012) with their medium resolution and broad band spectral coverage, $\sim 0.1-50\kev$ and $\sim 3-40\kev$ respectively, played crucial role in identifying and confirming the presence of UFOs along with \xmm{}. See Table 1 for a list of current and future missions and the charateristics of their spectrographs. Here we discuss the phenomenology and observations of these ionized outflows. Figure \ref{Fig:outflows} summarizes the physical parameters of the outflows.

\subsection{The broad and narrow UV absorption lines}

The AGN UV absorption lines are empirically classified into three classes depending on their widths 1. the broad absorption line (BALs), 2. the narrow absorption lines (NALs) and 3. the intermediate ``mini-BALs". The BALs are blue shifted relative to the AGN emission lines, implying outflow velocities in the range $v\sim 10000-60000\kms$, athough there is considerable diversity in their widths, a representative width being $\sim 10,000 \kms$. To qualify as BAL, the minimum requirement is to have velocity full width at half maximum (FWHM) $>3000\kms$ and blueshifted velocity $>5000\kms$ \citep{1981ARA&A..19...41W}. The NALs, on the other hand, typically have blue-shifts $<4000 \kms$ \citep{2017ApJ...848...79C}, although there are exceptional cases with higher velocities \citep[$\sim10,000 \kms$, ][]{2011MNRAS.410.1957H}. The intermediate class of ``mini-BALs" have properties in between NALs and BALs \citep{2004ASPC..311..203H}.

\begin{figure*}
  \centering

	\hbox{
	\includegraphics[width=9.5cm,angle=0]{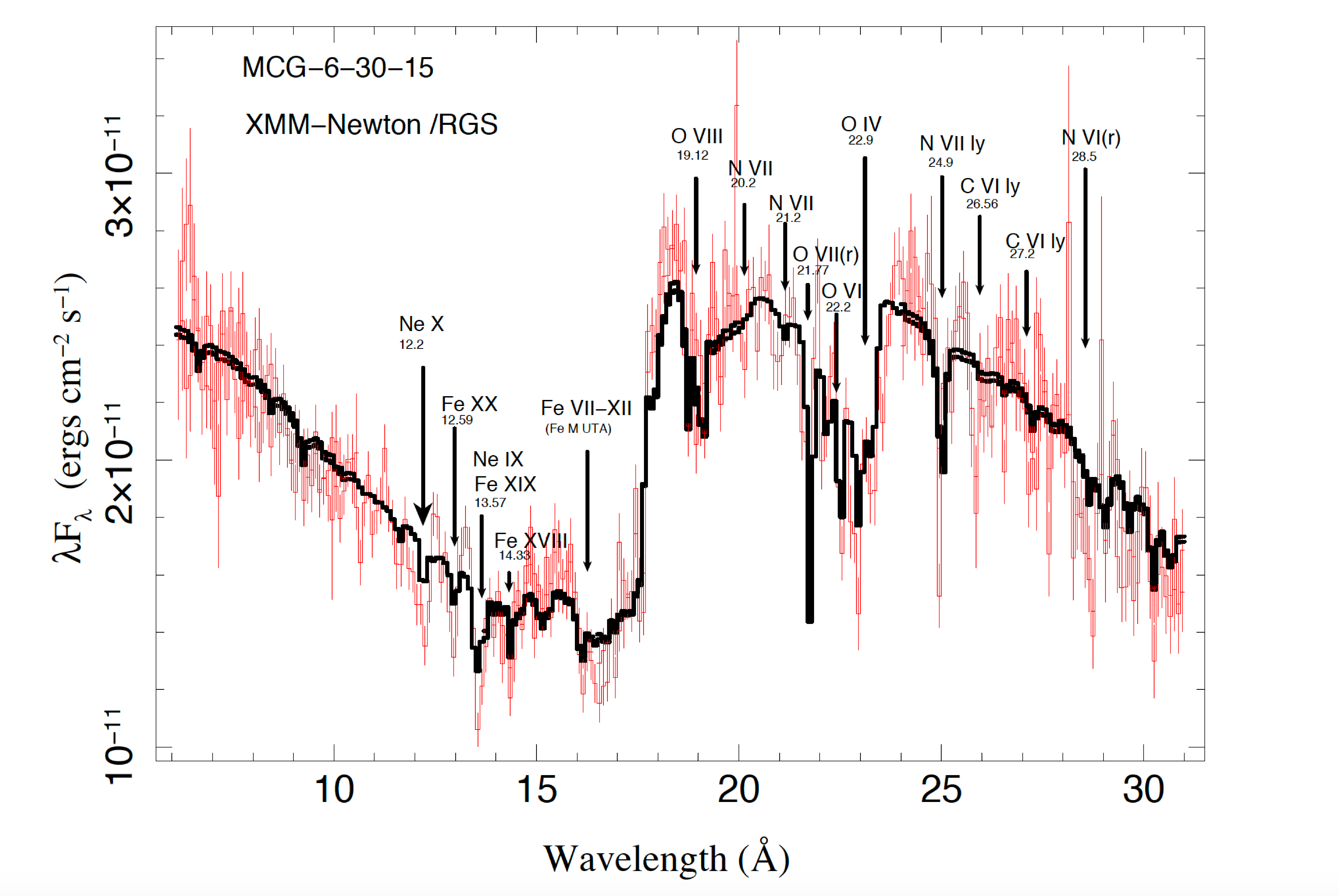}
	\includegraphics[width=9.5cm,angle=0]{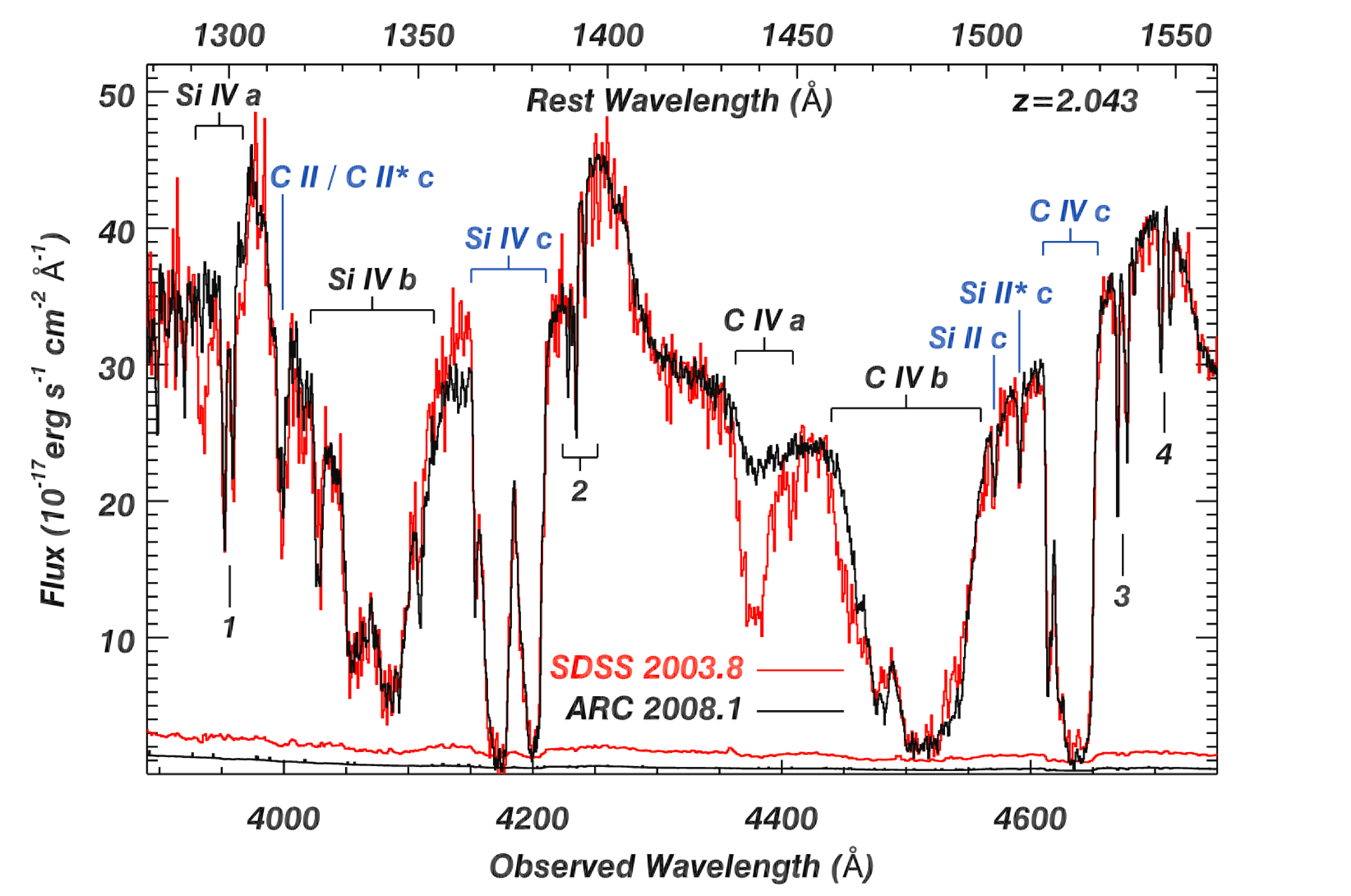}
	}
	\caption{{\it Left:} The warm absorbers in the source  MCG-6-30-15 detected using \xmm{} reflection grating spectrometer \citep{2014MNRAS.441.2613L}. {\it Right:} The BAL troughs in SDSS J0838+2955 detected using the Astrophysical Research Consortium 3.5 m telescope \citep{2009ApJ...706..525M}.} \label{Fig:WA}
\end{figure*}

Nearly $\sim 15-20\%$ of the observed quasars exhibit signatures of broad blueshifted absorption troughs (BALs) in UV spectra \citep{1981ARA&A..19...41W,2009ApJ...692..758G} detected using transitions from ions such as Si IV $\lambda 1400$, C IV $\lambda 1549$, Al III $\lambda 1857$ and Mg II $\lambda 2799$ \citep[see for e.g.,][ and the references therein]{2003AJ....125.1784H,2008MNRAS.386.1426K,2009ApJ...692..758G}, with relativistic speeds $\sim 0.1-0.2c$ \citep{2013MNRAS.435..133H}. See Fig \ref{Fig:WA} right panel for BAL absorption features in a quasar. Theoretical modeling has clearly shown the kinematic importance of feedback by these outflows \citep{2010ApJ...722..642O,2010ApJ...717..708C,2010MNRAS.406..822M,2010MNRAS.401....7H,2012MNRAS.425..605F,2014ApJ...789...19H}. In addition, the BAL outflows play an important role in shaping the observed quasar spectral properties in different wavelength bands. For example, the BALs are mostly associated with 1) high ionization emission lines, 2) optical/UV reddening and 3) X-ray absorption \citep{1981ARA&A..19...41W,2004ApJ...611..125L,2011AJ....141..167R,2009ApJ...692..758G}. The BAL QSOs have more reddened UV spectra than non-BAL QSOs implying large absorption columns of gas and dust, which points towards a radiation driven flow for these massive outflows \citep{2002ApJ...569..641L}. It is believed that an X-ray absorber shields the BAL outflow from over-ionization by the soft X-rays from the AGN, so that the UV radiation can accelerate it to high velocities \citep{1995ApJ...451..498M}, and sometimes the BALs may also be self-shielded \citep{2000ApJ...543..686P}. Interestingly it has also been found that BAL quasars are X-ray weak compared to non-BAL quasars, and sources with higher UV luminosity have higher BAL velocity \citep{2009ApJ...692..758G}.

BAL troughs have been known to vary on timescales of years or shorter, implying that most BAL absorption is formed within an order of magnitude of the wind-launching radius, although some of the BAL troughs may arise on larger scales \citep{2007ApJ...656...73L,2013ApJ...777..168F,2018MNRAS.481.5570V}. As in the case for all outflows detected in absorption, measuring the distance $r$ from the central black hole is extremely difficult. Typically in the case of BALs, one uses absorption line ratios from singly ionized ions such as Fe II and Si II to obtain the density and hence the distance using the definition of the ionization parameter $\xi$. But recent sample studies of BALs \citep{2018ApJ...857...60A} using higher ionization absorption lines ({\sc{S}{IV}}) found that $50\%$ of the quasar outflows are at distances larger than $100\pc$ and at least $12\%$ at distances $>1\kpc$. These results make it difficult to distinguish between mechanisms that may be driving the winds. Radiative acceleration can be effective both near the disk and at larger radii \citep[e.g.,][]{1995ApJ...454L.105M,2000ApJ...545...63E,2000ApJ...543..686P}, but the strong magnetic fields required by MHD processes require an origin near the disk \citep[e.g.,][]{1982MNRAS.199..883B,1994ApJ...434..446K,2004ApJ...615L..13E,2010ApJ...715..636F,2018ApJ...864L..27F}. However, the winds seen at large radii may simply reflect the terminal velocity of a disk-launched wind. How and where the BAL outflows are formed have important implications for the energetics of the outflows in these systems.

Interestingly, BALs and other high velocity absorption line profiles occur predominantly in the higher luminosity Quasars and rarely in lower-luminosity AGN such as Seyfert galaxies, and BALs also tend to be absent when there is a strong radio emission, while the NALs are found with radio emission \citep{2008MNRAS.386.1426K,2019MNRAS.483.1808H}. These may tell us about the unique physical interaction between the AGN and the different types of outflows (radiative mode and kinetic mode).

In some cases, blue-shifted narrow absorption lines at redshifts very close to the emission redshift of the AGN (associated systems) are detected \citep{1999ApJ...516..750C,2000ApJ...535...58K,2003ARA&A..41..117C,2003A&A...403..473K,2011A&A...534A..41K}. These outflows are sometimes important in mass transfer, but due to their small velocity, they are mostly insignificant contributors to AGN feedback. In this context it is worth mentioning the narrow emission lines which in some nearby AGN could be spatially resolved and hence their distance and mass-outflow rates can be estimated \citep{2018ApJ...856...46R,2015ApJ...799...83C,2018ApJ...867...88R}. For e.g., studies with \hst{} and {\it Gemini} GMOS spectrograph focussing on [O III] emission lines in sources such as NGC~4151 \citep{2005AJ....130..945D} revealed radial outflows with velocity structures, and ionized accelerated knots, the mass outflow rate peaked at $\mout\sim 3.0 \msolyi $ at a distance of $\sim 70\pc$ from the central SMBH, indicating that most of the energy of the outflow is deposited within $100\pc$ from the SMBH, and the outflows perish beyond that distance scale. These also point out that these low velocity systems may not be effective in feedback.

In some cases however, the narrow HI Ly$\alpha$ absorption lines detected in UV have high blueshift indicating a large outflow velocity ($\sim 17,000\kms$). These may be interesting cases of UFO being detected in the UV \citep[for e.g., in the source PG~1211+143][]{2018ApJ...853..166K,2018ApJ...853..165D}. A systematic search for UV counterpart of UFOs (at the expected location of HI Ly$\alpha$ absorption) in a sample of AGN which host UFOs has however, lead to a null result \citep{2018ApJ...859...94K}.


\subsection{The warm absorbers}

The warm absorbers are detected as absorption lines and edges from H-like and He-like ions of C, O, N, Ne, Mg, Al, Si and S, in the soft X-ray spectra (see Fig \ref{Fig:WA}  left panel) of $\sim 65\%$ of the nearby AGN \citep{1997MNRAS.286..513R,2005A&A...431..111B,2007MNRAS.379.1359M,2014MNRAS.441.2613L}. These absorbers are always found blueshifted with respect to the systemic redshift, implying that these are always in outflow  ($v\sim 100-2000\kms$) and never in inflow. In a few cases where the signal to noise (SNR) and resolution are high enough to warrant detections of individual unblended absorption lines, it has been found that the lines are asymmetric having extended blue wings \citep{2002ApJ...574..643K}. Apart from absorption lines and edges there are deep absorption troughs detected in the rest frame wavelength $15-17\rm \AAA$, possibly arising out of blended Fe M shell unresolved transition arrays (UTA). The absorbers producing these features have a low ionization parameter \citep[$\rm log\xi\sim 0-1$, ][]{2001A&A...365L.168S} and carry a significant amount of mass outflow rate \citep{2001A&A...365L.168S,2016MNRAS.457.3896L}. However, as their velocity is low $\sim 100-1000\kms$ the kinetic luminosity $\Ek$ may not be sufficient for feedback. Nevertheless, these are important mechanisms by which the SMBH can deposit matter near its surroundings.

In most cases, the WA clouds have been found to thrive in a multiphase medium with a range in $\xi$, $\nh$ and even in $v$, \citep{2004ApJ...611...68K,2017A&A...601A..17B,2014ApJ...793...61H,2013MNRAS.435.3028E,2009ApJ...690..773K,2016A&A...587A.129E} and very rarely detected as a single cloud \citep[see for e.g., theoretical works by ][]{2009MNRAS.393...83C}. To quantify this effect, an important parameter is the absorption measure distribution (AMD). The AMD is a measure of the column density of the cloud as a function of its ionization parameter (AMD$={\rm d}\log\nh/{\rm d}\log\xi$). The shape of the AMD depends heavily on the irradiating SED and the density of the warm absorber, thereby serving as crucial diagnostics of the plasma as well as AGN-WA relations  \citep{2019ApJ...881...78A}. In most cases the AMD has been found to be multi-peaked, a testimony of the existence of multi-phase ionic absorbers along the line of sight, with unstable ionization phases in the gaps of the distribution \citep{2007ApJ...663..799H}, roughly corresponding to the ionization states of $0.5<\log(\xi/\xiunit)<1.5$. Interestingly, such ionization gaps have been detected in sample studies of local AGN \citep{2007MNRAS.379.1359M,2014MNRAS.441.2613L}. The measure of AMD also gives us the density profile of the outflows ($n_{\rm e}\propto r^{-\alpha}$), where $\alpha$ is the radial gradient of the density and for WA the values are $\alpha=1.236\pm 0.034$ \citep{2013MNRAS.430.1102T,2016MNRAS.457.3896L}. The density profile is a crucial parameter to estimate the mass outflow rate ($\mout$) and kinetic luminosity ($\Ek$) of the outflows. 


 WA response to changing AGN continuum luminosity and shape holds the key to crucial information on the origin and acceleration mechanisms of WA. For example., the variations in the ionization state of the WA in response to the AGN continuum and flux changes can yield the recombination timescale $t_{\rm R}$, which is the time required by the free electrons to recombine to a bound energy level of an ion. $t_{\rm R}$  is inversely proportional to the cloud electron density \citep{1999ApJ...512..184N}, and an estimate of the cloud electron density ($n_{\rm e}$) gives us the distance of the cloud ($r$) from the central source from the definition of $\xi$. Several works using high resolution spectra in X-rays have successfully used this technique to measure the density and hence the distance. The range of density estimated is $n_{\rm e}\sim 10^{4}-10^{11}\cmcubei$, indicating a distance of a few $10$s of $\pc$ down to $\le 0.01\pc$ \citep{2012ApJ...746L..20K,2007ApJ...659.1022K,2012A&A...539A.117K,2014ApJ...793...61H,2003ApJ...599..933N,2013ApJ...776...99R}.  Long term monitoring campaigns of well known variable sources such as NGC~4051, NGC~3783, NGC~5548, Mkn~509, have been undertaken in the last two decades to use this method. For example, the soft X-ray spectra of the source Mrk~509 revealed five WA components with \xmm{}, and the distance of the three components could be put around $\sim 5-400\pc$, where the slow outflow components point to an origin in the NLR or torus region \citep{2012A&A...539A.117K}. Similarly, the opacity variations in X-ray WA of NGC~3783 revealed deep absorption troughs due to Fe M-shell absorbers and the structure of the absorber is heavily clumped, disfavoring a continuous radial range of ionization structures \citep{2005ApJ...622..842K}.

\begin{figure*}
  \centering

	\hbox{
	\includegraphics[width=8cm,angle=0]{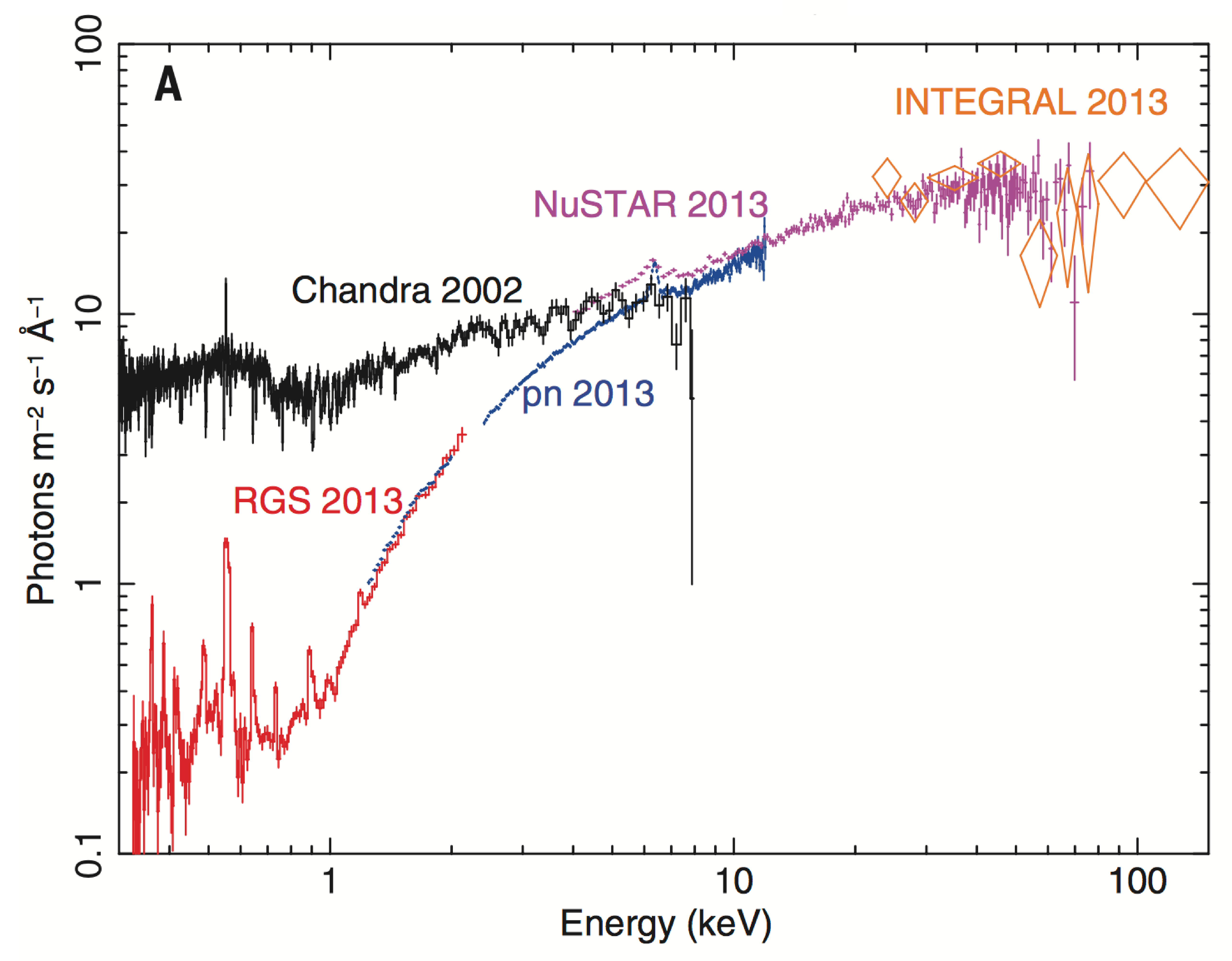}
	\includegraphics[width=8.5cm,angle=0]{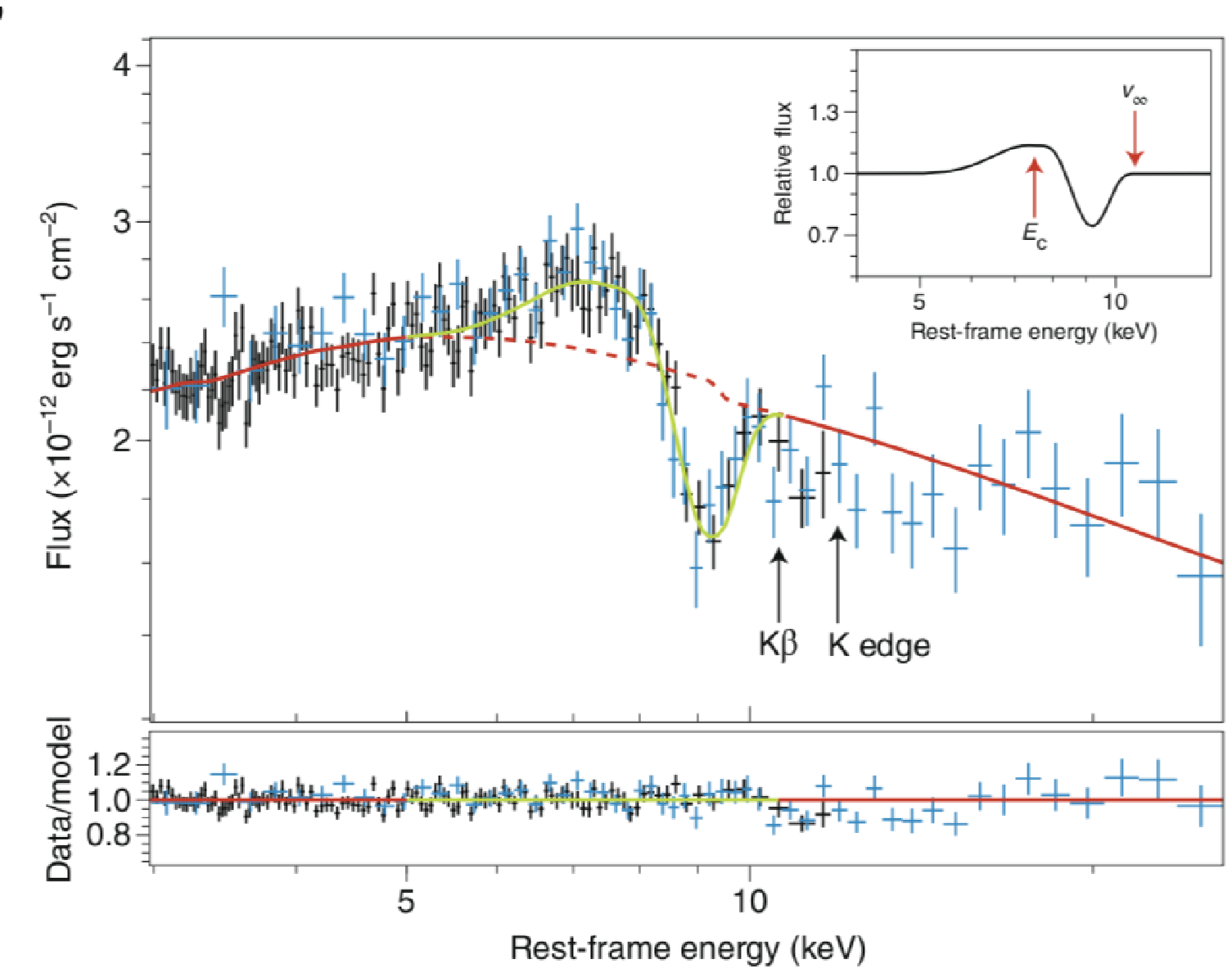}
	}
	\caption{{\it Left:} The transient X-ray obscurers in NGC~5548 causing simultaneous soft X-ray and UV obscuration \citep{2014Sci...345...64K}. {\it Right:} The UFO detected as blue-shifted absorption lines in the quasar PDS~456, along with a PCygni profile in the FeK band \citep{2015Sci...347..860N}.} \label{Fig:pcygni}
\end{figure*}

The recent discovery of a transient, low ionized ($\log\xi\sim 1.8$), but high column density ($\log\nh\sim 23$) X-ray obscurer during a monitoring campaign of the source NGC~5548 has revealed a new type of X-ray obscurer \citep[See Figure \ref{Fig:pcygni} left panel, ][]{2014Sci...345...64K}. It has been suggested that a long-lasting clumpy stream of ionized gas passed across our line of sight causing simultaneous soft X-ray and UV absorption troughs. The obscurer blocks nearly $90\%$ of the soft X-rays. The outflow velocity of the gas was noted to be around five times faster than those in the persistent outflow, and are located at a distance of only a few $\sim$ light-days from the central SMBH, indicative of an accretion disk wind reaching beyond BLR. These outflows are best explained by magnetic acceleration rather than by radiative acceleration, the latter leading to more radial and equatorial flow. The UV counterpart of the obscuring flow of NGC~5548 exhibited components of broad, blue-shifted absorption associated with Ly$\alpha$, N V, Si IV, and C IV in the \hst{} COS spectra \citep{2019ApJ...881..153K}. Although we still do not know the exact cause of such an obscuration flow, it has been suggested that a collapse of the broad line region has led to the outburst and triggered the obscuring event. There has been a few other incidences of this event in other sources such as Mrk 335 \citep{2013ApJ...766..104L}, NGC 985 \citep{2016A&A...586A..72E}, and NGC~3783 \citet{2017A&A...607A..28M}. Due to high column density, the X-ray spectral features of these obscuration events do not have any resolved features in the soft X-rays, and hence the UV (where individual absorption lines could be identified) provides measurement of the kinematics and ionization state of these absorbers.

The lower ionization states of the WA clouds are likely to contain dust. The dust component greatly enhances the effective thrust due to the radiation pressure, somewhere between a factor of $1-500$ depending on column density, and they may become a critical component in feedback momentum and kinematics from the central SMBH \citep{2008MNRAS.385L..43F,2018MNRAS.476..512I}. However, identifying the presence of dust using X-ray spectra is not straight forward. Earlier attempts used X-ray and IR spectra to estimate the dust embedded in WA \citep[see for e.g.,][]{1997MNRAS.291..403R,2001ApJ...554L..13L}. In several sources, the implied dust column density obtained from X-rays agreed with the reddening measurements obtained using IR studies. To date however, only a few such attempts with multi-wavelength studies have been made \citep[see for e.g., ][ for quantifying dust in absorbers for IC~4329A using spatialy resolved IR observations]{2018A&A...619A..20M}.

Warm absorbers are rarely found in the radio loud galaxies, much similar to the phenomenon that BAL outflows are detected mostly in radio quiet QSOs. WA outflows have been detected in radio-loud galaxies such as 3C 390.3, 3C 120, 3C 382, 3C 445 \citep{2009ApJ...702L.187R,2010ApJ...725..803R,2010MNRAS.401L..10T,2012MNRAS.419..321T,2016ApJ...830...98T}. This paucity in occurrence of WA in radio loud AGN may have deep implications on when the AGN creates an outflow and when it produces a jet (different modes of feedback). However, a direct physical relationship has not been established yet. Studies with samples of radio loud Seyfert-1 AGN detected a significant inverse correlation between the column density $\nh$ of the WA and the radio loudness parameter (R) of the jet \citep{2019A&A...625A..25M}. The relation could be explained from the perspective of a change in the magnetic field configuration (toroidal to poloidal), powering either the jet or the wind mode of outflow. Hence, when the jet is stronger, the WA is weaker and vice-versa. Interestingly, such anti-correlations between wind and jet properties have been seen in stellar mass black hole binaries \citep{2009Natur.458..481N,2012MNRAS.422L..11P}.


\subsection{The ultra-fast outflows}

The radically improved sensitivity and spectral resolution in the $2-10\kev$ energy range of \chandra{}, \xmm{} and \suzaku{} over previous missions have opened up the avenue for detection of highly ionized winds from the SMBH observed as Fe XXV and Fe XXVI K-shell absorption lines in the region $7-9\kev$. These absorbers may sometimes have velocities as high as $\sim 30\%$ of the speed of light and calculations suggest that they carry considerable kinetic power required for feedback by the AGN i.e, $\Ek >0.5\%\lbol$  \citep[][]{2010MNRAS.401....7H,2011ApJ...742...44T,2015MNRAS.451.4169G}. Perhaps as much as $40\%$ of nearby AGN show the presence of UFO \citep{2010A&A...521A..57T,2011ApJ...742...44T,2013MNRAS.430...60G}, with the outflow velocity spanning from $\sim 0.03-0.3 c$, ionization parameter $\log(\xi / \xiunit)\sim 3-6$, and column density $\log(\nh/\cmsqi)\sim 22-24$. 

Understanding how these winds are formed, in particular, the interplay between the accretion and ejection flows at small radii in the accretion disk can give us vital clues on the energetic output of these outflows and hence the feedback to galaxies. The origin of the UFOs is still debated, but radiatively driven flows may not be an efficient driver in these cases as the UFOs are higly ionized and do not have much opacity in UV or X-ray lines \citep{2014ApJ...789...19H}. Magneto hydrodynamical effects in the accretion disks (magnetic fields threading the disks) may play a vital role in generating these fast outflows \citep{1982MNRAS.199..883B,2012ASPC..460..181K,2010ApJ...715..636F,2017NatAs...1E..62F,2018ApJ...852...35K}. The radial distance of these outflows estimated using the virial velocity is $10^{2-4}r_{\rm s}$, where $r_{\rm s}$ is the Schwarzschild radius of the SMBH, indicating proximity to the accretion disk \citep{2015MNRAS.451.4169G}. Strong correlation between the bolometric luminoisty ($\lbol$) of the AGN and the kinetic luminosity ($\Ek$) of the UFO have been detected, indicating that the faster winds are more likely to be driven in higher luminosity or high Eddington AGN, however, they may not necessarily be radiation driven \citep{2015MNRAS.451.4169G}. While its possible that Eddington limited AGN may produce fast, radiatively driven winds, more generally the correlation could arise as a result of the increasing dominance of the inner accretion disk (and its MHD effects) down to small radii in more luminous sources \citep{2013MNRAS.430.1102T}. The wind velocity $v$ has been found to be separately significantly correlated with $\lbol$ with a slope $\beta=0.4_{-0.2}^{+0.3}$. Similarly the momentum flux of the radiation ($\pbol$) and that of the wind ($\pwind$) are strongly corretaled with a slope of $\beta=1.2_{-0.7}^{+0.8}$.

There have been suggestions that the UFOs may originate at the base of the radio jet \citep{2004A&A...413..535G}, but, the connection between radio jets and the UFOs is not solidly established yet. By comparing samples of radio loud and radio quiet AGN, it has been found that the UFO detection fraction in both these cases are similar ($\sim 50\pm 20\%$), implying that the presence of relativistic jets does not preclude the existence of these winds \citep{2014MNRAS.443.2154T}.

 The opening angle of the UFOs is a vital parameter to estimate the  total kinetic power carried by the winds and hence a measure of the feedback, but the detection of the absorption lines alone does not provide us this information. The PCygni profiles of FeK emission and absorption lines detected in a few bright quasars gives us a rare view of the opening angle of the UFOs \citep{2015Sci...347..860N,2019ApJ...884...80R}. The principle here is that the UFO plasma absorbs (along the line of sight) and also re-emits (isotropic) and hence if we detect both emission and absorption from the same plasma, the relative strength can give us an estimate of the opening angle. For example, the PCygni profiles detected in the \xmm{} and \nustar{} observations of the quasar PDS~456 yielded a solid angle (filled by the winds) $\Omega=3.2\pm0.6\pi$ sr (See Figure \ref{Fig:pcygni} right panel).

The high z quasars are an important laboratory for studying the AGN phenomenon, as these are the periods (redshift $z\sim 1-3$) when the quasar luminosity function peaked. It is commonly believed that the SMBH `negative' feedback may have produced quasar downsizing, and hence studying feedback through outflows in the quasars at these redshifts are important to understand the luminosity function evolution. Early studies by \citet{2002ApJ...579..169C} detected high velocity Fe XXVI absorption features in the lensed BAL quasar APM 08279+5255 at a redshift of $z=3.91$. \citet{2018A&A...610L..13D} and \citet{2015ApJ...799...82C} detected similar such high velocity outflows in high redshift quasars. However, high redshift studies to date remain only a few and the next generation instruments such as \athena{} and \lynx{} can shed more light on these interesting sources.

Apart from the so called traditional UFOs detected in the $6-9\kev$ as Fe absorption features, there have been detections of high velocity outflows using high resolution grating spectra in the soft X-rays ($0.3-2\kev$), such as in PG~1211+143 \citep{2003MNRAS.346.1025P,2016AN....337..518P,2018ApJ...854...28R}, IRAS~17020+4544 \citep{2015ApJ...813L..39L}, Mrk~590 \citep{2015ApJ...798....4G}, Ark~564 \citep{2013ApJ...772...66G} and PDS~456 \citep{2020ApJ...895...37R}. The UFOs detected in the soft X-ray spectra have velocities in the range  $\sim10,000 \kms-0.2c$, but with a much lower column density ($\log(\nh/\cmsqi)\sim 22$) and the ionization parameter ($\log(\xi/\xiunit)\sim 3$) compared to that of the traditional UFOs detected in the FeK band.

 The response of the UFOs to varying AGN continuum flux serves as an important indicator of how and when these outflows are generated. Studies on the variable quasar PDS~456 with \xmm{}, \nustar{} and \suzaku{} observations found that the centroid energy of the blue shifted FeK absorption profile increases with AGN luminosity, and the wind outflow velocity correlates with the hard X-ray luminosity as $v/c\propto L_{7-30\kev}^{0.22\pm0.04}$, which indicates that the wind is predominantly radiatively driven due to the high Eddington ratio of the source  \citep{2017MNRAS.472L..15M}. Such occurences have also been found in other sources accreting at high Eddington ratio indicating that the UFOs may be driven by intense radiation pressure in sources accreting at or close to the Eddington rate, such as in APM 08279+5255 \citep{2003ApJ...595...85C}, PG~1211+143 \citep{2003MNRAS.345..705P}, IRAS F11119+3257 \citep{2015Natur.519..436T}, 1H 0707-495 \citep{2016MNRAS.461.3954H}, and IRAS~13224-3809 \citep{2017MNRAS.469.1553P}. Interesting anti-correlations with the $3-10\kev$ luminosity and the UFO absorption equivalent width have been detected for the source IRAS~13224-3809 \citep{2017Natur.543...83P} indicating a connection between the accretion process and the outflow generation.

Although rare, there has been a few tentative detections of ultra-fast ``inflows" indicative of feeding of the black hole accretion disk. These features are detected as absorption at $\sim 5.5\kev$ and under the assumption that they arise from H- or He- like Fe, implied infall velocities are estimated to be $\sim 0.15-0.2c$ \citep[see for e.g.,][detected in Mrk~509 using {\it BeppoSAX}]{2005A&A...442..461D}. More recent findings of transient absorption signatures of highly ionized Fe lines in the $\sim 5-6\kev$ pointed towards a short lived  ultra-fast inflow with an inflow velocity of $v\sim 0.3c$ in nearby quasars \citep[PG~1211+143, ][]{2005ApJ...633L..81R,2018MNRAS.481.1832P}.

It is important to note that UFOs have also been found in stellar mass blackhole candidates such as IGR J17091-3624 \citep{2012ApJ...746L..20K}. Photoionization modeling of these winds point towards an origin within 43,300 Schwarzschild radii of the black hole and the BH may be expelling more gas than it accretes. These are interesting case studies because the stellar black hole candidates are possibly the scaled down versions of AGN \citep{2006Natur.444..730M}.

Although UFOs are of paramount importance in terms of kinematics and SMBH feedback, it has only been possible to detect these outflows in a few dozen AGN, mostly due to lack of SNR and/or spectral resolution in these energy bands. Future telescopes {\it XRISM} and {\it Athena} will be able to characterize these outflows in much greater detail (See Section \ref{Sec:instruments}).

\begin{figure*}
  \centering

	\hbox{		
	
	\includegraphics[width=8cm,angle=0]{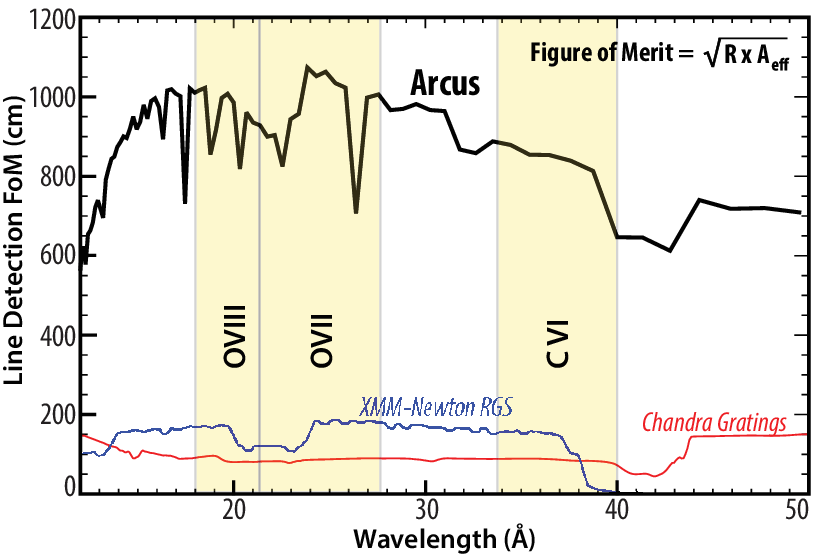}
	\includegraphics[width=7cm,angle=0]{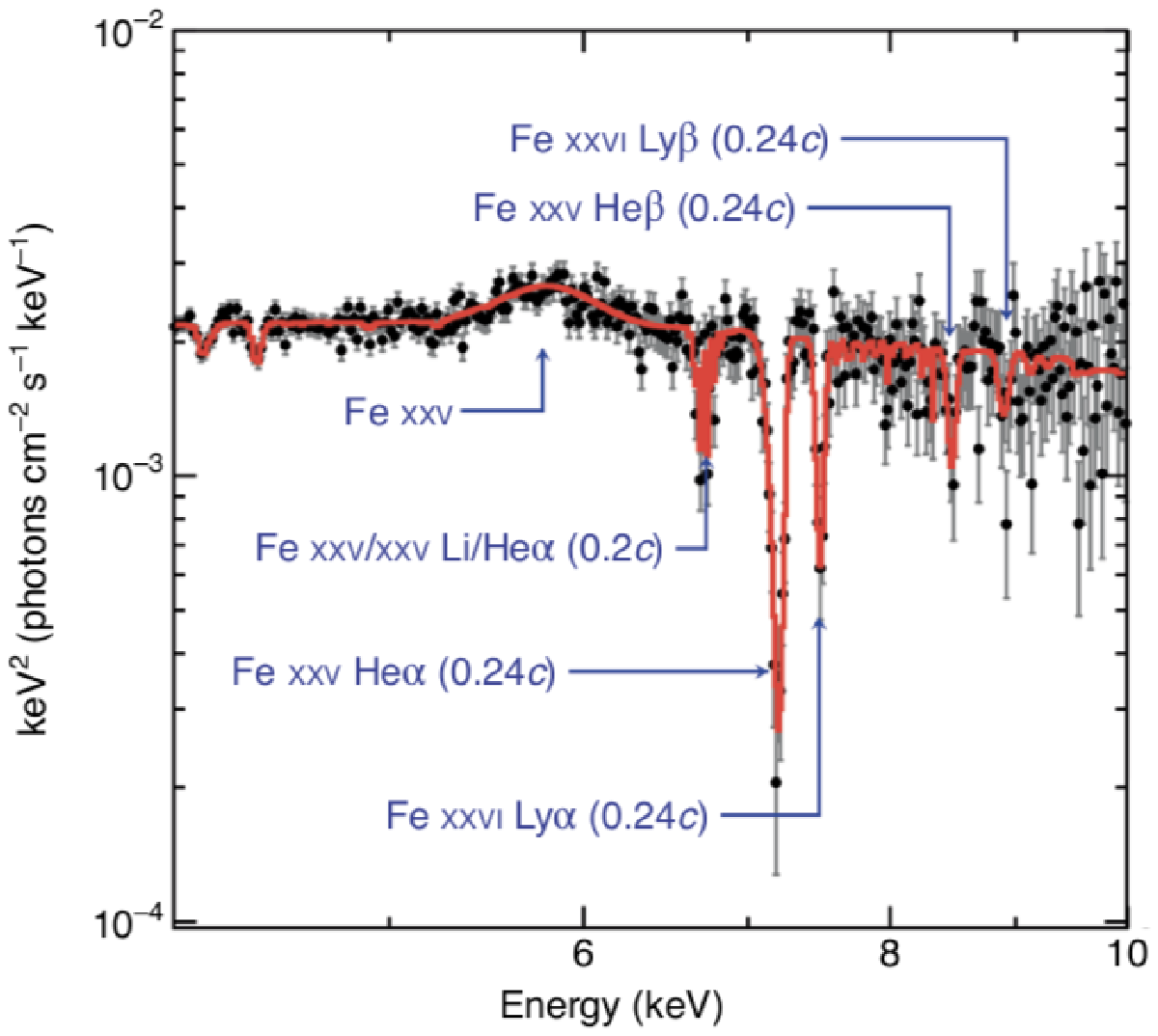}
	
	}
	\caption{{\it Left} The enormous improvement in the figure of merit by the proposed mission {\it Arcus} over the current missions \xmm{} and \chandra{} in the crucial soft X-ray energy band where the density and temperature diagnostic emission and absorption lines are detected. {\it Right:} The high resolution detection of UFOs by integral X-ray field unit (X-IFU) on board the future mission \athena{}. {Picture courtesy: Arcus and Athena team. }} \label{Fig:Arcus}
\end{figure*}

\subsection{The connection between the outflows} \label{subsec:connection}

Figure \ref{Fig:outflows} shows the typical parameter ranges of the different ionized outflows detected in AGN. We note that there is considerable overlap in the distance scales as well as the outflow parameters ($\xi$, $\nh$ and $v$) between the NALs and the WA. Similar such overlap is found in the UFO and the BAL outlfows. However, a direct link between these outflows has not yet been established, due to several limiting factors, such as, different sensitivity of the instruments in the energy bands where these outflows are detected, lack of accurate distance estimates of these outflows etc. Keeping in mind these caveats, we can possibly say that the UFO and the BALs originate from the innermost regions of the accretion disk, and the BALs may be the lowest ionization tails of UFOs. Future work on a systematic search for BALs in systems where lower ionization UFOs (soft X-ray) are detected can provide more insights. A few BAL quasars have also shown UFO \citep[for example, Mrk~231, APM~08279+5255][]{2015A&A...583A..99F,2002ApJ...579..169C}. However, it is not clear if they form the same outflow, or the UFO drives the BAL. 

  Attempts to connect the UV NAL outflows and the warm absorbers as a part of a single phase wind \citep{1994ApJ...434..493M} have met with mixed response for different sources. Recent multiwavelength high resolution campaigns on sources such as NGC~3783, NGC~5548, NGC~4593 have found that the soft-X-ray outflows are kinematically similar to the UV outflows, but most of the WA material is in a too high ionization state to produce measurable lines in the UV spectra \citep{2015A&A...577A..37A,2000ApJ...535L..17K,2001ApJ...554..216K,2003ApJ...583..178G,2001ApJ...551..671K,2003ApJ...597..832K}.

   The correlations of the plasma parameters of the UFOs and the WA point towards a possibility that they are a part of the same stratified outflow, the UFOs originating much nearer to the SMBH compared to the WA \citep{2013MNRAS.430.1102T}. Supporting this idea, it has been conjectured that the warm absorbers may be generated via a Compton-cooled shocked wind, when a UFO launched very near to the SMBH, loses most of its mechanical energy after shocking against the interstellar medium \citep{2013MNRAS.433.1369P}. For example in the narrow line Seyfert 1, NGC~4051, with a black hole mass of $\log(\mbh/\msol) \sim 6$, the shock radius turns out to be $\sim 10^{17}\cm$, much in line with sample studies on WA and UFO \citep{2013MNRAS.430.1102T,2016MNRAS.457.3896L}. In some sources UFOs are detected simultaneously in the soft X-rays and also in the FeK band. Such detections have led to a view that these absorbers may have a multi-phase origin, as if a clumpy-ISM has been entrained by an inner UFO, resulting in a simultaneous flow of both the WA and UFO \citep{2019A&A...627A.121S}. All these evidences lead us to ask if WA outflows are the `missing link' between the UFOs and the molecular outflows.

  AGN are believed to be one of the main drivers behind the kpc scale molecular outflows \citep{2013ApJ...776...27V,2014A&A...562A..21C,2017A&A...601A.143F,2018ApJ...868...10L}. These large scale molecular outflows carry large amount of kinetic luminosity and mass
outflow rate and are capable of removing materials from the host galaxy and also regulate the star
formation rate. However, the exact mechanism by which the AGN interacts with the molecular gas
in the host galaxy and drives the outflows is not known. X-ray outflows are a very important
candidate for such an interaction, either by energy or momentum conserving processes \citep[see for e.g.,][]{2014A&A...562A..21C,2015Natur.519..436T,2017ApJ...843...18V,2020A&ARv..28....2V}. IRAS F11119+3257 is one example where the molecular outflow (traced by neutral NaI, OH absorption and CO emission lines) have been found to be driven by an inner UFO in an energy conserving way \citep{2015Natur.519..436T,2017ApJ...843...18V}. Similarly, the BAL quasar Mrk~231 where both UFO and molecular outflows have been detected \citep[as traced by CO(2-1) and CO(3-2) emission lines][]{2015A&A...583A..99F}, it is found that the UFO is driving the molecular outflow (MO) in an energy conserving way. On the other hand for the source PDS~456 an energy conserving mechanism of the UFO creating the MO may be ruled out \citep{2019A&A...628A.118B}. In the ultra luminous infra-red galaxy (ULIRG) IRAS~F05189-2524, it has been found that the UFO drives the molecular outflow in an energy conserving way, although the momentum driven mechanism may not be completely ruled out \citep{2019ApJ...887...69S}.


\section{The origin and acceleration mechanisms of the ionized outflows}\label{Sec:origin}

Depending on their ionizations states, different outflows can have different origin and acceleration mechanisms \citep{2003ARA&A..41..117C}. As we discuss the scenarios below, we must note that the different driving mechanisms have different implications on how the feedback works, and hence have impact on the bigger questions of black hole growth and black hole-galaxy co-evolution. In the last couple of decades simulations of accretion disks and outflows have progressed rapidly and clearly demonstrate that thermal, magnetic, radiation, and shock driven forces, or a combination of these are able to generate and accelerate winds, resulting in different end results (in terms of feedback and matter movement). It is therefore crucial to identify in what circumstances certain mechanisms dominate and understand their impact on black hole growth and co-evolution.

\begin{itemize}

	\item \textbf{MHD winds:}  It is suspected that for most AGN outflows in general, MHD processes may be at work in some form and intensity, because most important AGN nuclear phenomenon like the accretion disc (viscosity), the existence of corona, and jets can be explained by MHD mechanisms \citep{1982MNRAS.199..883B,1994ApJ...434..446K,2004ApJ...615L..13E,2010ApJ...715..636F,2018ApJ...864L..27F}. An MHD scenario is naturally invoked to explain the high velocity of UFOs which are highly ionized outflows and hence have virtually no line driven force due to the incident radiation. In case of a MHD driven wind, simple scaling relations, such as $\vout \propto \xi$ may hold true \citep[see for e.g.,][]{2012ASPC..460..181K}. The MHD scenario seems to explain well the disk winds in the source NGC~3783 \citep{2018ApJ...853...40F}, where the soft X-ray warm absorption features in  the wavelength range $1.8-20 \rm \AAA$ could be well described by an MHD model with a cloud density profile of $n(r)\propto r^{-1.15}$ and inclination angle of the disk $\theta=44^{\circ}$. Similar results have been found for the WA and UFO absorbers detected in the source PG~1211+143 \citep{2015ApJ...805...17F}. The MHD winds have been detected in several LMXBs, where the magnetic processes are not only involved in driving the outflows but also for mediating mass transfer within the disk itself \citep[see for e.g.,][and references therein]{2019ApJ...886..104T,2006Natur.441..953M,2008ApJ...680.1359M,2015ApJ...814...87M,2017NatAs...1E..62F,2012ApJ...750...27N}

	\item \textbf{Radiative winds:} Radiation pressure due to UV and X-ray absorption lines are important mechanisms for driving the outflows where the cloud opacity is high enough \citep{2004ApJ...616..688P}. For highly ionized clouds the UV and X-ray line opacities may not be sufficient enough to drive them, and Compton scattering can provide the necessary momentum to the outflow, particularly for AGN radiating near/at Eddington rate \citep{2003MNRAS.345..657K,2010MNRAS.402.1516K}. An important quantity in the radiatively driven flows is the force multiplier M($\xi$, $t$) which measures the thrust by radiation on the clouds. Theoretical studies of radiatively driven outflows in variable sources \citep[see for e.g., NGC~5548][]{2019ApJ...882...99D} found that M($\xi$, $t$) starts to decrease for $\xi\ge 1$, and the decrease is gradual and can be non-monotonic as M($\xi$, $t$) can increase by a factor of a few in the range $\xi=10-1000$. This increase of M at high $\xi$ is for very low optical depths, so even if line driving is important for $\xi$ as high as 1000, the outflow may produce only very weak features that are hard to detect in absorption.

		Recent studies have pointed out that most of the diversity detected in the AGN accretion/ejection flows, in the context of radiatively driven winds, can be described well by varying just two fundamental parameters of the SMBH: $\mbh$ and $\lambdaedd$ \citep{2019A&A...630A..94G}. It has been conjectured that physical models of line-driven disc winds can open up an avenue to estimate black hole growth \citep{2018arXiv181101966N}, because there is possibly a direct connection between the mass accretion rate of the SMBH and wind power (accretion ejection connection).

		Motivated by the discoveries of UFOs in PG1211+143 \citep{2003MNRAS.345..705P} with velocity $\sim 0.1c$ and column density $\nh\sim 10^{24}\cmsqi$, suggesting that the clouds could have been Compton-thick at smaller radii, it has been proposed that black holes accreting at or above the Eddington rate probably produce optically thick winds and drive them by radiative pressure, both in the cases of SMBHs and LMXBs \citep{2003MNRAS.345..657K}. For the particular case of PG1211+143 it was found that the mass outflow rates of the winds $\mout\sim 1.6\msolyi$ was comparable to the Eddington accretion rate, implying a radiatively driven accretion disk wind. Recent 2D hydrodynamical simulations have also considered the re-radiation photons (locally produced radiation by Thompson scattering and Bremsstrahlung radiation), which resulted in a better match with observed properties of the outflows \citep{2019MNRAS.490.2567M}.

	An interesting study on thermal stability of winds driven by radiation pressure in ultra-luminous X-ray sources (ULXs) \citep{2020MNRAS.491.5702P} found that at super-Eddington luminosity, the radiation pressure is expected to inflate the accretion disk and drive fast winds (fraction of the speed of light), and these winds are in stable thermal equilibrium. These studies confirm that radiation pressure is capable of driving fast winds.

\item \textbf{Thermal winds:} X-rays from the AGN heat up the disk material to the Compton temperature $T_{\rm IC}$. The heated layer expands due to the pressure gradient, eventually producing a thermally driven wind at radii where the sound speed is larger than the local escape velocity \citep{1983ApJ...271...70B,1996ApJ...461..767W,2001ApJ...561..684K,2002ApJ...565..455P,2006ApJ...652L.117N,2008ApJ...687...97D,2010ApJ...719..515L}. The thermal wind models predict that the column density of the wind is roughly proportional to the mass accretion rate of the central SMBH \citep{2019ApJ...884..111K,2018MNRAS.473..838D}. Owing to its slow velocity and mild level of ionization, several works have suggested a thermal origin of the WA, having a large launch radius extending beyond BLR and/or torus \citep{1995ApJ...447..512K,2001ApJ...561..684K,2008ApJ...687...97D,2019MNRAS.489.1152M}. Thermal winds are highly sensitive to the irradiating SED, and that serves as an important factor in constraining the parameters \citep{2017MNRAS.467.4161D}. The possibility of thermally driven winds have been investigated in stellar mass black hole binaries and NS binary systems and have been found to explain the outflows well \citep[see for e.g.,][]{2018MNRAS.473..838D,2019MNRAS.490.3098T}, or sometimes in a hybrid form incorporating both MHD  and thermal winds \citep{2018MNRAS.481.2628W}. The idea of having hybrid acceleration mechanisms active has also been extended to AGN \citep{2008ApJ...687...97D,2019MNRAS.489.1152M}, where the Compton temperature of the wind has been tracked as a function of $\mbh$ and $\lambdaedd$, and a combination of thermal and radiation pressure acting on dust have been suggested to drive the WA outflows.

\end{itemize}

\section{The open questions.}\label{Sec:outstanding_questions}

Here we list some of the most important outstanding questions in AGN feedback physics, the answers to which hold the key
to our comprehensive understanding of the nature and growth of SMBH and galaxies.

\begin{itemize}

	\item What are the origins of the outflows, the acceleration mechanisms, and how the outflowing gas couples with the host galaxy ISM? Extensive variability studies of the absorbers will help us to construct a comprehensive physical, spatial and temporal picture of the AGN winds.  Several observations have remained unexplained in the present theoretical perspectives. For example, what drives the anti-correlation between broad absorption lines features and radio emission in QSOs? Why are the BAL quasars X-ray weak?, Why is there an anti-correlation between radio loudness of an AGN and WA column density, while there is no such signatures for UFOs? What is the real cause behind the phenomenon of rapid variability of X-ray obscurers observed in sources such as NGC~5548 \citep{2014Sci...345...64K}? What is the detection rate of ionized outflows in radio loud AGN?

	\item  Do these ionized outflows always exist in multi-phase medium. Are these multi-phase media signature of a gradient of plasma properties along the line of sight, that is a continuous flow?  Do these multi-phase media exist in equilibrium with each other?

	\item  The central engines of the AGNs are commonly believed to be scaled up versions of black hole X-ray binaries \citep{2006Natur.444..730M} and hence it will be interesting to compare the physical properties of the ionized outflows from binaries with that of the AGNs. Do we find similarity in the outflow physics at these two extreme ends of black hole mass ranges? We note that the SEDs are very different in the two cases.

	\item  Are AGN ionized outflows the main drivers of the large scale molecular outflows? It is still not clearly known how the ionized outflows (UFOs) couple with the host galaxy gas and dust to drive these massive outflows. Current estimates \citep[e.g.,][]{2017A&A...601A.143F} show a wide range of momentum boost factors connecting the large-scale molecular gas to the inner, fast X-ray winds and the handful that have been co-observed \citep[e.g.,][]{2019A&A...628A.118B} do not exclude either energy or momentum conserving feedback in general. Further sub-mm observations of confirmed fast X-ray winds, coupled with refined kinematical measurements of the latter through calorimeter-resolution instrumentation (see Section 5), may clarify this picture in forthcoming years.

	\item  The  accretion-ejection connection is an open question which may hold the key to some of the profound AGN questions: When does the SMBH stop feeding? How does the accretion support the ejection flows (disk winds)? etc. The indirect signatures of accretion-ejection connection is established in a sample study of nearby AGN \citep{2017Natur.549..488R} where it is found that the fraction of obscured AGN drops sharply with increase in the source Eddington rate ($\lambdaedd$), implying that the line of sight obscuration (which may be feeding the SMBH) is blown off by the AGN itself beyond certain levels of accretion.

\end{itemize}


\section{Future perspectives}\label{Sec:instruments}

Addressing these fundamental open questions requires that we develop next generation telescopes with significantly improved spectral resolution as well as increased effective area, while maintaining or improving existing spatial resolution. The current high spectral resolution optical/UV and X-ray missions have demonstrated their power in unfolding the nature of outflows in an unprecedented way, and now we need to take the next steps. These missions, of course, will also require improvements in theoretical modeling and atomic databases to match the new observations.

\subsection{Instrumentation}

Existing X-ray spectrometers have useful resolutions well under $R\le1000$, strongly limiting the ability to resolve features and connect them to material seen in other bandpasses. Observatories with higher spectral resolution ($R>2000$ at $1\kev$) will resolve individual absorption lines and study the asymmetries and shifts in the line profiles revealing important information about outflow structures and their impact. This must be paired with the higher effective areas ($\ge 1000 \rm cm^2$; compared to existing missions with areas of $1-100 \rm cm^2$ required to study the outflows in more distant quasars, particularly at the quasar peak era (redshift $1 < z < 3$) when the AGN population was most luminous. 

These next generation X-ray telescopes with high spectral resolution and high throughput will not merely extend existing data, they will unveil properties and impact of highly energetic X-ray outflows currently inaccessible.  With $R=2000$ at $1\kev$ the velocities of X-ray absorbing clouds can be measured to $<20 \kms$, along with their column density. Combined with the density diagnostic metastable and ground state transitions the true mass in these clouds can be accurately determined.

New missions will also allow studies on large enough sample of sources to derive statistically significant information. A simultaneous high resolution UV + X-ray mission will encompass the crucial AGN ionizing continuum ($13.6\ev$-$100\kev$), and also characterize the simultaneous detections of UV and X-ray outflows, which map different spatial scales along the line of sight. Planned missions such as XRISM and Athena, and concept studies such as Arcus, Lynx and LUVOIR hold the key to our understanding of the outflows from AGN. Figure \ref{Fig:Arcus} left and right panels demonstrate the superior capabilities in spectral resolution and effective area of the missions \arcus{} and \athena{} respectively, compared to the present generation X-ray missions.

\subsection{Theoretical Modelling}

Our prospects of understanding the origin, acceleration mechanism and the impact of AGN winds will crutially depend on future `realistic' theoretical modeling of the spectral features in the UV, X-rays, and other energy bands. Until recently, most interpretations of
observations in the context of the wind model has been done by
comparison with spectra computed for one-dimensional constant-density
slabs of photoionized gas, commonly obtained from codes 
such as XSTAR and CLOUDY \citep{2001ApJS..133..221K,2017RMxAA..53..385F}. This does not fit into the realistic picture of an outflowing wind because the kinematic structure of the flow and its time evolution need to be considered in detail. More complex theoretical models involving these effects need to be generated to actually interpret the data. Several recent works have focused on these aspects using the self-consistent calculation of the hydrodynamics of plasma irradiated by AGN SED \citep[][]{2017MNRAS.467.3160W,2017ApJ...836...42H,2020ApJ...893L..34D}. Certain preliminary results point out that dramatic changes in the flow pattern does not necessary translate to significant changes in the predicted spectra \citep[see for example, ][]{2007A&A...474....1M,2017MNRAS.467.3160W}, however, more work needs to be done. Similarly, the special relativistic effects in simulating AGN outflows need to be considered \citep{2010MNRAS.408.1396S,2020A&A...633A..55L}.

\subsection{Atomic database}

The basic atomic data involving strong transitions, charge state distributions in photoionization and collisional equilibrium have been reasonably well measured to date.  However, the new instruments and theoretical studies will require models of time-dependent plasmas, as well as improved models of weak (but numerous) features such as dielectronic satellite lines whose wavelengths and fluxes have been shown to be quite uncertain \citep{2019ApJ...871...50B,2018PASJ...70...12H}. Additionally, the developers of atomic spectral codes will need to develop approaches to deal with uncertainties or errors in the underlying atomic data, as the current approach of assuming zero systematic errors due to the spectral modeling are now clearly inadequate \citep{2018PASJ...70...12H}.

\begin{table}
{\footnotesize
	\caption{The instrument specifications for current and next generation telescopes. }\label{Table:instruments}
	  \begin{tabular}{lllllllllllllll} \hline\hline 
		  Mission		&Instrument	&Energy Range	&Spectral resolution		& 			&  \\
					&		&		&				&			 \\ \hline \\
	   
		  \xmm{}	&RGS 	  	& $0.4-2\kev$	&$\sim 500$				&\\ 
		  \chandra{}	&HETGS	  	& $0.4-10\kev$	&$\sim 800$				&\\ 
		  \hst{}	&STIS   	& $1150-10300\rm \AAA$	& $750-100,000$			&	\\ 
		  \hst{}	&COS(FUV)	& $900-3200\rm \AAA$    & $1500 - 18000$			&\\ \hline
	 
		  \athena	&X-IFU		&$0.2-12\kev$	& $2800$					&	\\ 
		  \arcus	&--		&$12-50\rm \AAA$	&$3000$						&\\  
		  \xrism	&Resolve	&$0.3-12\kev$ & $60-2400$					&	\\  
		  \lynx		&X-ray grating	&$0.2-2\kev$	&$5000$						&	\\  
		  \luvoir	&LUMOS		&$1000-25000\rm \AAA$ & $500-65,000$				&	\\ \hline \\

\end{tabular}
Note that the spectral resolution is a function of energy. The numbers quoted here are typical ranges for a given instrument operating in an energy interval.

}
\end{table}


\clearpage
\bibliographystyle{mn2e}

\bibliography{mybib}

\end{document}